\documentclass[
   final            % use final for the camera ready runs
 ,nofootinbib                % add further options here if necessary
  ]{aipproc}

\layoutstyle{8x11single}

\usepackage{graphicx}
\usepackage{amssymb}
\usepackage{amsmath}

\def\a{\alpha}
\def\b{\beta}

\def\p{\phi_a}
\def\ptil{\tilde\phi_a}

\newcommand{\maton}{\texttt{maton}}

\newcommand{\susyflavor}{\texttt{susy\_flavor}}

\begin{document}
\begin{flushright}
OHSTPY-HEP-T-13-004
\end{flushright}

\title{SO(10) Yukawa Unification after the First Run of the LHC}


\author{Stuart Raby}{
  address={Department of Physics, The Ohio State University, 191 W. Woodruff Ave., Columbus, OH 43210, USA}
}

\begin{abstract}
In this talk we discuss SO(10) Yukawa unification and its ramifications for phenomenology.  The initial constraints come from fitting the
top, bottom and tau masses, requiring large $\tan\beta \sim 50$ and particular values for soft SUSY breaking parameters.  We perform a global $\chi^2$ analysis, fitting the recently observed `Higgs' with mass of order 125 GeV in addition to fermion masses and mixing angles and several flavor violating observables.   We discuss two distinct GUT scale boundary conditions for soft SUSY breaking masses.  In both cases we have a universal cubic scalar parameter, $A_0$.  In the first case we consider universal gaugino masses, and universal scalar masses, $m_{16}$, for squarks and sleptons;  while in the latter case we have non-universal gaugino masses and either universal scalar masses, $m_{16}$, for squarks and sleptons or D-term splitting of scalar masses.  We discuss the spectrum of SUSY particle masses and consequences for the LHC.
\end{abstract}

\maketitle

\section{Introduction}

Fermion masses and mixing angles are manifestly hierarchical.   The simplest way to describe this hierarchy is with Yukawa matrices which
are also hierarchical.   Moreover the most natural way to obtain the hierarchy is in terms of effective higher dimension operators of the form
\begin{equation}  W \supset 16_3 \ 10 \ 16_3 + 16_3 \ 10 \ \frac{45}{M} \ 16_2 + \cdots.
\end{equation}
This version of SO(10) models has the nice features that it only requires small representations of SO(10),  has many predictions
and can, in principle, find an UV completion in string theory.  There are a long list of papers by authors such as Albright, Anderson, Babu,
Barr, Barbieri, Berezhiani, Blazek, Carena, Chang, Dermisek, Dimopoulos, Hall, Masiero, Murayama, Pati, Raby, Romanino, Rossi, Starkman,
Wagner, Wilczek, Wiesenfeldt, and Willenbrock which have followed this line of model building.

The only renormalizable term in $W$ is $\lambda \ 16_3 \ 10 \ 16_3$ which gives Yukawa coupling unification
\begin{equation}  \lambda = \lambda_t = \lambda_b = \lambda_\tau = \lambda_{\nu_\tau}  \end{equation} at $M_{GUT}$.
Note,  one CANNOT predict the top mass due to large SUSY threshold corrections to the bottom and tau masses, as shown in
\cite{Hall:1993gn,Carena:1994bv,Blazek:1995nv}.  These corrections are of the form
\begin{equation}  \delta m_b/m_b  \propto \frac{\alpha_3 \ \mu \ M_{\tilde g} \ \tan\beta}{m_{\tilde b}^2} +
\frac{\lambda_t^2 \ \mu \ A_t \ \tan\beta}{m_{\tilde t}^2} + {\rm log \ corrections} .
\end{equation} So instead  we use  Yukawa unification to predict the soft SUSY breaking masses!!  In order to fit the data,
we need \begin{equation} \delta m_b/m_b \sim - 2\% . \end{equation}   For a short list of references on this subject, see \cite{Blazek:2001sb,Blazek:2002ta,Baer:2001yy,Auto:2003ys,Tobe:2003bc,Dermisek:2003vn,Dermisek:2005sw,Baer:2008jn,Baer:2008xc,Baer:2009ie,Badziak:2011wm,Gogoladze:2011aa,Anandakrishnan:2012tj,Anandakrishnan:2013cwa,Anandakrishnan:2013nca}.

\section{Gauge and Yukawa Unification with Universal Gaugino Masses}
\begin{figure}[h]
\includegraphics[width=20pc]{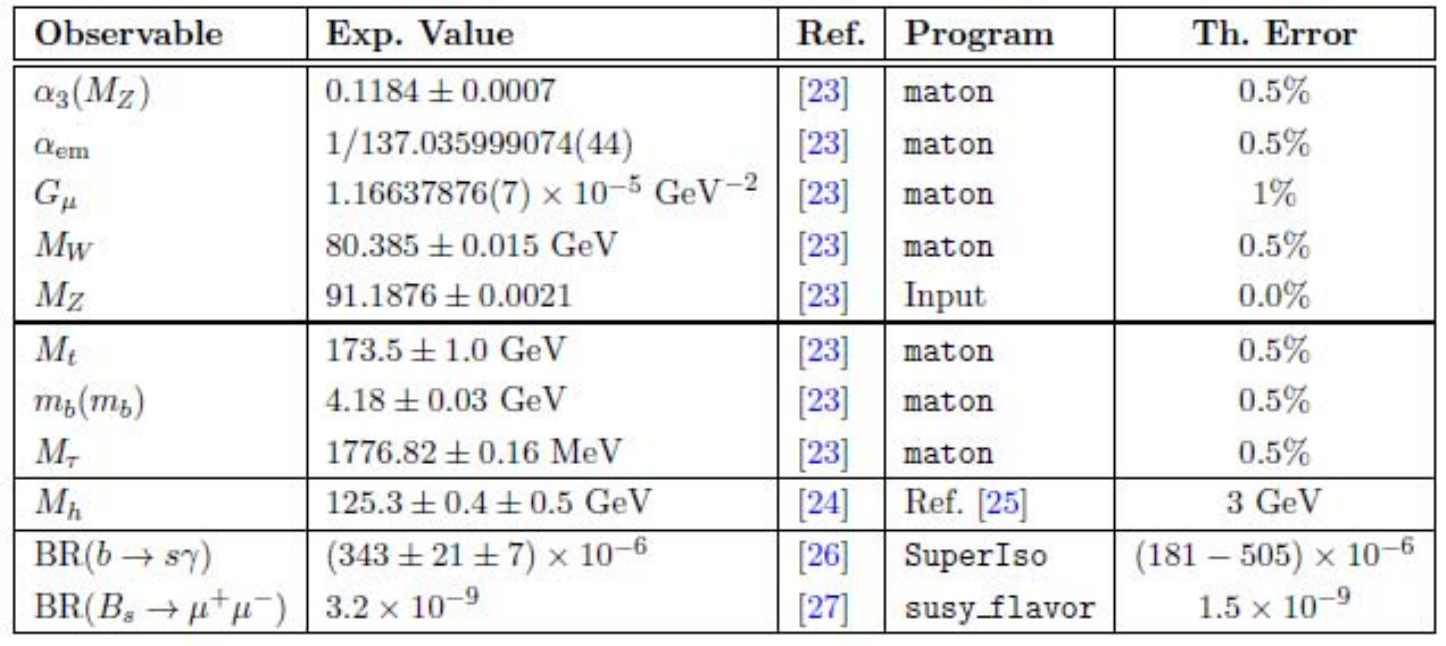}\hspace{2pc}%
\includegraphics[width=15pc]{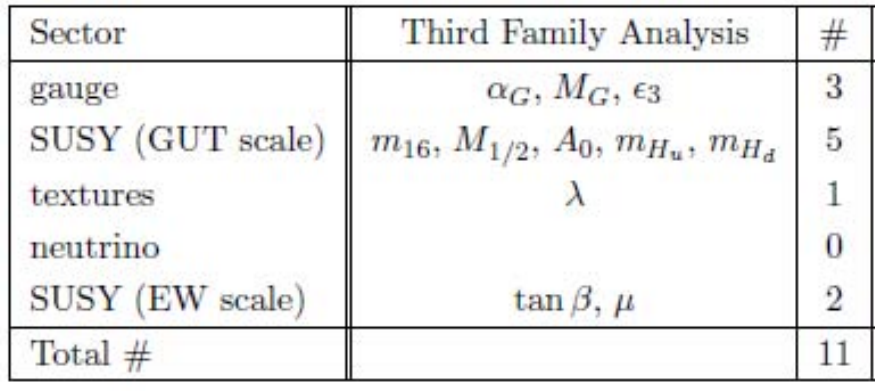}\hspace{2pc}%
\begin{minipage}[b]{14pc}\caption{\label{observables}(Left) The 11 low energy observables which enter the $\chi^2$ function. (Right)
The 11 arbitrary parameters which are varied to minimize $\chi^2$.}
\end{minipage}
\end{figure}

In the first case we take $\mu \ M_{\tilde g} > 0$, thus we need $\mu \ A_t < 0$ \cite{Anandakrishnan:2012tj,Anandakrishnan:2013nca}.
We assume the following minimal set of GUT scale boundary conditions --  universal squark and slepton masses,  $m_{16}$,  universal cubic scalar parameter, $A_0$, universal gaugino masses, $M_{1/2}$, and non-universal Higgs masses [NUHM] or `just so' Higgs splitting,  $m_{H_u}, \ m_{H_d}$  or
$m_{H_{u (d)}}^2 = m_{10}^2 [ 1 - (+) \Delta_{m_H}^2 ]$.  We then perform a global $\chi^2$ analysis fitting the 11 observables
as a function of the 11 arbitrary parameters,  Fig. \ref{observables}.
We find that fitting the top, bottom and tau mass forces us into the region of SUSY breaking parameter space with
\begin{equation}   A_0 \approx  - 2 m_{16},   \;\;  m_{10} \approx  \sqrt{2} \ m_{16}, \;\;  m_{16} > \ {\rm few \ TeV}, \;\;  \mu, M_{1/2} \ll m_{16}; \end{equation}
and, finally,  \begin{equation} \tan\beta \approx 50 . \end{equation}  In addition, radiative electroweak symmetry breaking requires
$\Delta_{m_H}^2 \approx 13\%$, with roughly half of this coming naturally from the renormalization group running of neutrino Yukawa couplings
from $M_G$ to $M_{N_\tau} \sim 10^{13}$ GeV \cite{Blazek:2001sb,Blazek:2002ta}.

It is very interesting that the above region in SUSY parameter space results in an inverted scalar mass hierarchy at the weak scale with the third family scalars significantly lighter than the first two families \cite{Bagger:1999sy}.  This has the nice property of suppressing flavor changing
neutral current and CP violating processes.

\subsection{Heavy squarks and sleptons}

\begin{figure}[h]
\includegraphics[width=14pc]{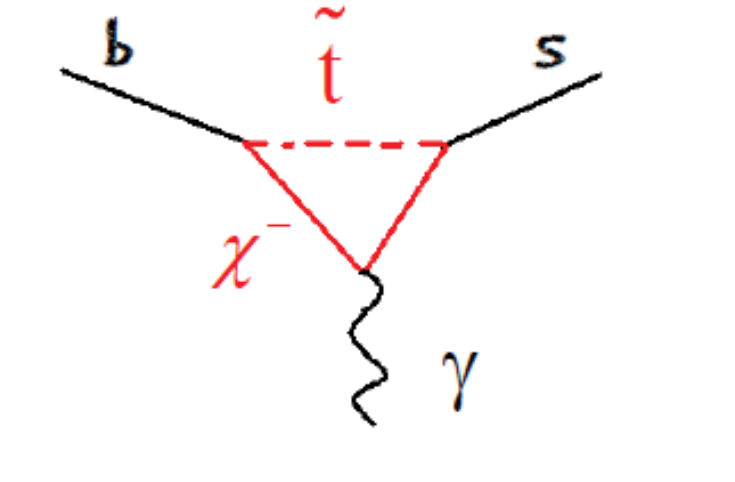}\hspace{2pc}%
\begin{minipage}[b]{14pc}\caption{\label{charginoloop}Dominant contribution to the process $b \rightarrow s \ \gamma$ in the MSSM.}
\end{minipage}
\end{figure}

Considering the theoretical and experimental results for the branching ratio $BR(B \rightarrow X_s \gamma)$, we argue that $m_{16} \ge 8$ TeV.   The experimental
value $BR(B \rightarrow X_s \gamma)_{\rm exp} = (3.55 \pm 0.26) \times 10^{-4}$, while the NNLO Standard Model theoretical value is
$BR(B \rightarrow X_s \gamma)_{\rm th} = (3.15 \pm 0.23) \times 10^{-4}$.  The amplitude for the process $B \rightarrow X_s \gamma$ is proportional to
the Wilson coefficient, $C_7$.   $C_7 = C_7^{SM} + C_7^{SUSY}$ and, in order to fit the data, we see that $C_7 \approx \pm C_7^{SM}$.  Thus
$C_7^{SUSY} \approx - 2 C_7^{SM} \; {\rm or} \; C_7^{SUSY} \approx 0$.
The dominant SUSY contribution to the branching ratio comes from a stop - chargino loop with
$C_7^{SUSY} \sim C_7^{\chi^+} \sim \frac{\mu \ A_t}{\tilde m^2} \ \tan\beta \times sign(C_7^{SM})$ (see Fig. \ref{charginoloop}).  Hence, in the former case (which allows for light scalars) $C_7 \approx - C_7^{SM}$, while in the latter case (with heavy scalars) $C_7 \approx C_7^{SM}$.

\begin{figure}[h]
\includegraphics[width=14pc]{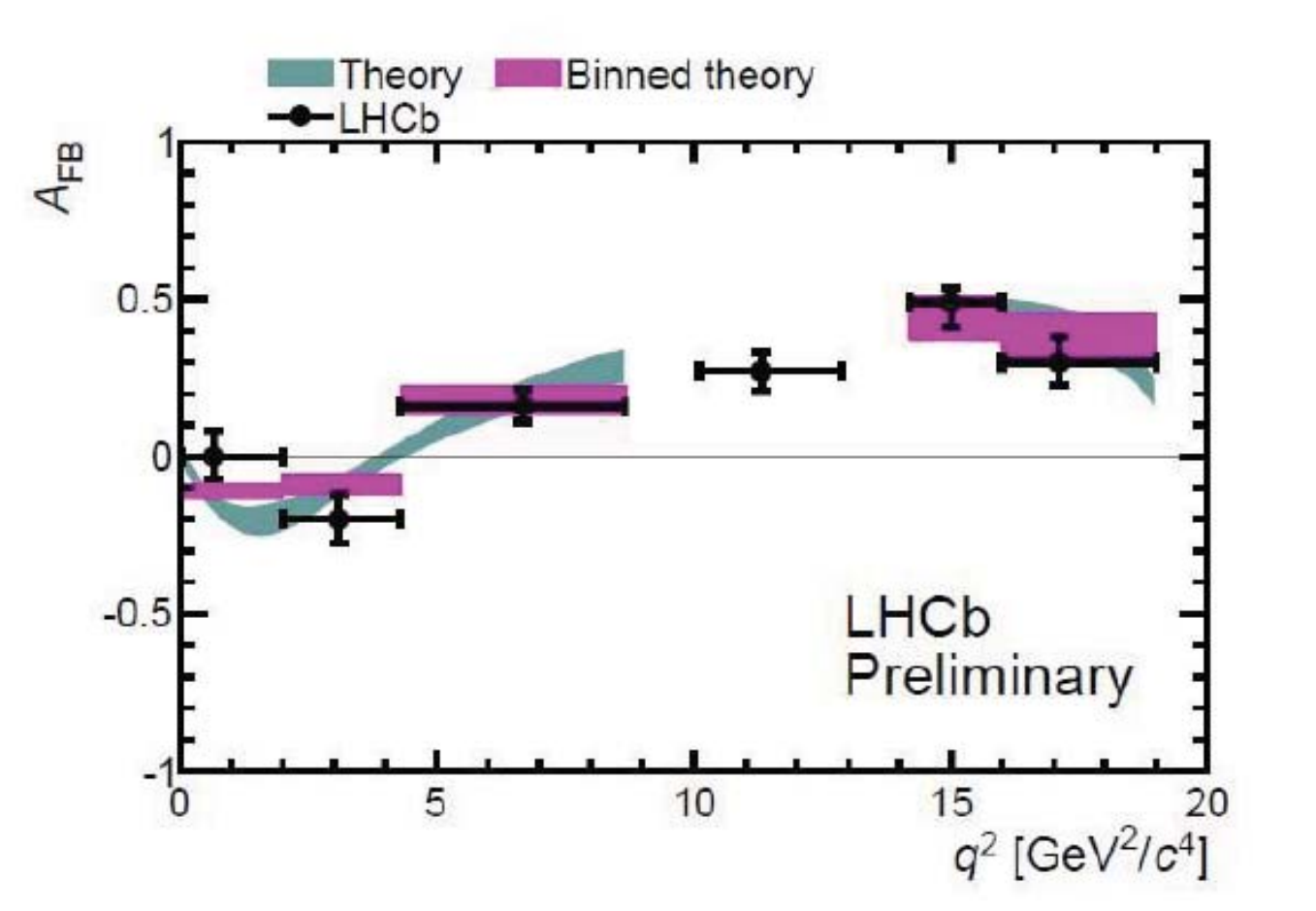}\hspace{2pc}%
\begin{minipage}[b]{14pc}\caption{\label{bkmumu}The forward-backward asymmetry for the process $B \rightarrow K^* \ \mu^+ \ \mu^-$ measured by LHCb.}
\end{minipage}
\end{figure}

Recent LHCb data on the $BR(B \rightarrow K^* \ \mu^+ \ \mu^-)$ now favors  $C_7 \approx + C_7^{SM}$ \cite{Bediaga:2012py} (see Fig. \ref{bkmumu}).
This tension between the processes  $b \rightarrow s \gamma$ and $b \rightarrow s \ \ell^+ \ \ell^-$ was already discussed by
Albrecht et al. \cite{Albrecht:2007ii}.  In order to be consistent with this
data one requires $C_7^{\chi^+} \approx 0$  or  $C_7 \approx C_7^{SM} + C_7^{SUSY} \approx + C_7^{SM}$ and therefore  $m_{16} \ge 8$ TeV.

\begin{figure}[h]
\includegraphics[width=30pc]{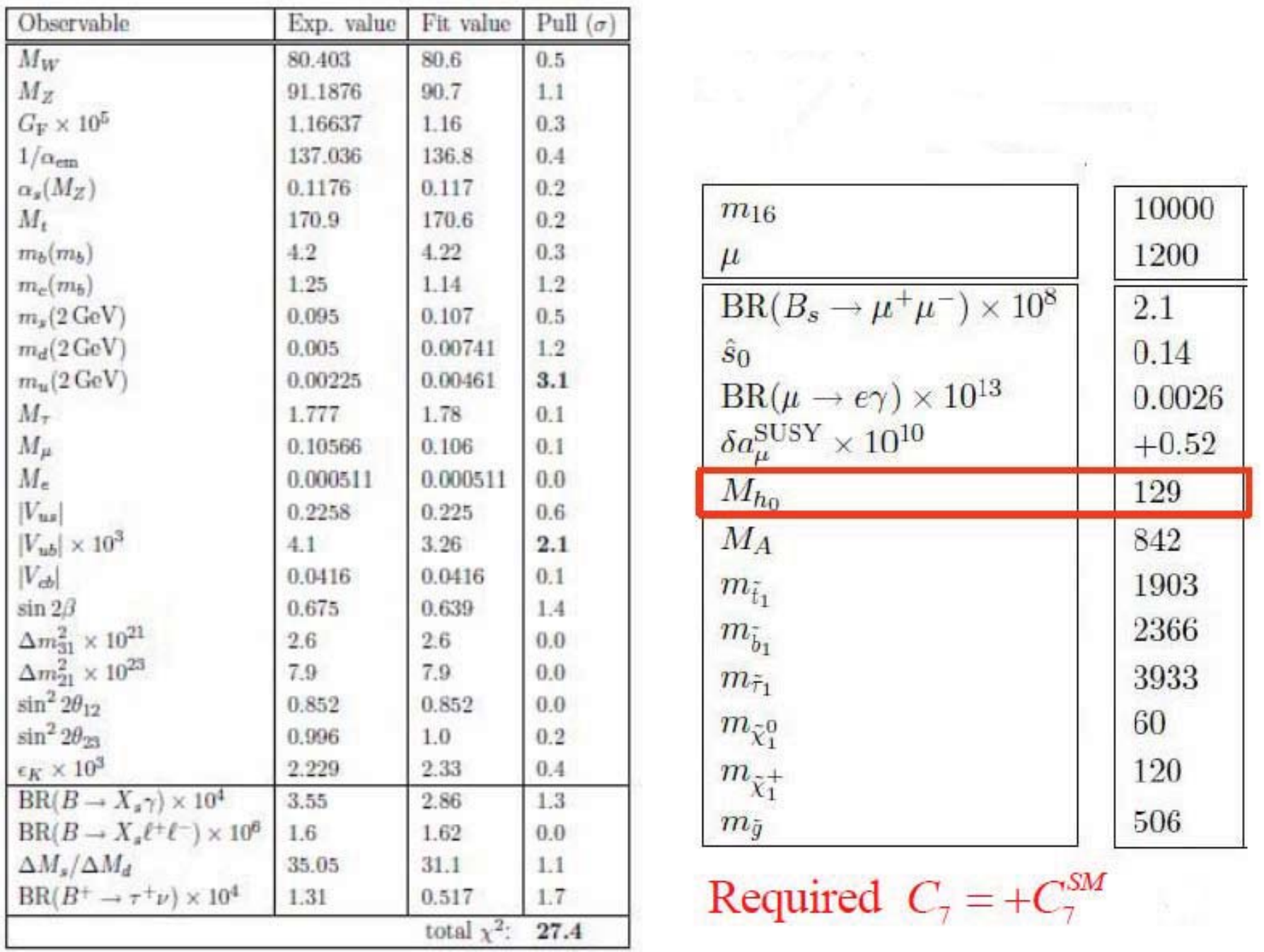}\hspace{2pc}%
\begin{minipage}[b]{30pc}\caption{\label{albrecht}The results obtained by Albrecht et al. for $m_{16} = 10$ TeV.}
\end{minipage}
\end{figure}

In 2007,  Albrecht et al. \cite{Albrecht:2007ii} performed a global $\chi^2$ analysis of this theory (including the Yukawa structure for all three families).  Two of the tables from their paper are exhibited in Fig. \ref{albrecht}. This analysis included 27 low energy observables and a reasonable fit to the data was only found for $m_{16} = 10$ TeV.   Note, the Higgs mass was predicted to be 129 GeV.

\subsection{Light Higgs mass}

An approximate formula for the light Higgs mass is given by \cite{Carena:1995wu}
\begin{equation} m_h^2 \approx M_Z^2 \ \cos^22\beta + \frac{3 g^2 m_t^4}{8 \pi^2 m_W^2} \left[\ln\left(\frac{M_{SUSY}^2}{m_t^2}\right) +
\frac{X_t^2}{M_{SUSY}^2} \left(1 - \frac{X_t^2}{12 M_{SUSY}^2}\right)\right] \end{equation}  where
$X_t = A_t - \mu/\tan\beta$.   The light Higgs mass is maximized as a function of $X_t$ for $X_t/M_{SUSY} = \pm \sqrt{6}$, referred to as maximal
mixing.   Hence we see that for large values of $A_t$ and $M_{SUSY}$ it is quite easy to obtain a light Higgs mass of order 125 GeV.

\subsection{$B_s \rightarrow \mu^+ \ \mu^-$}

In this section we argue that the light Higgs boson must be Standard Model-like.  To do this we show that the CP odd Higgs boson, $A$,
must have mass greater than $\sim$ 1 TeV and as a consequence this is also true for the CP even Higgs boson, $H$, and the charged Higgs bosons,  $H^\pm$, as well. This is the well-known decoupling limit in which the light Higgs boson couples to matter just like the Standard Model Higgs.

Consider the branching ratio $BR(B_s \rightarrow \mu^+ \ \mu^-)$ which in the Standard Model is  $\sim 3 \times 10^{-9}$.   In the MSSM this receives
a contribution proportional to $\sim  \frac{\tan\beta^6}{m_A^4}$.   Recent experimental results give \cite{:2012ct}
\begin{eqnarray}
LHCb \; : &  = (3.2 \stackrel{+1.5}{-1.2} \pm 0.2) \times 10^{-9} \; &  {\rm with} \; 1 \ {\rm fb}^{-1} (7 \; {\rm TeV}) \; {\rm and} \; 1.1 \ {\rm fb}^{-1} (8 \; {\rm TeV})  .  \end{eqnarray}    Since we have $\tan\beta \sim 50$, our only choice is to take the
CP odd Higgs mass to be large with $m_A \ge 1$ TeV.   This is the decoupling limit; {\em hence the light Higgs is SM-like.}

\subsection{Gluino Mass}
\begin{figure}[h]
\includegraphics[width=20pc]{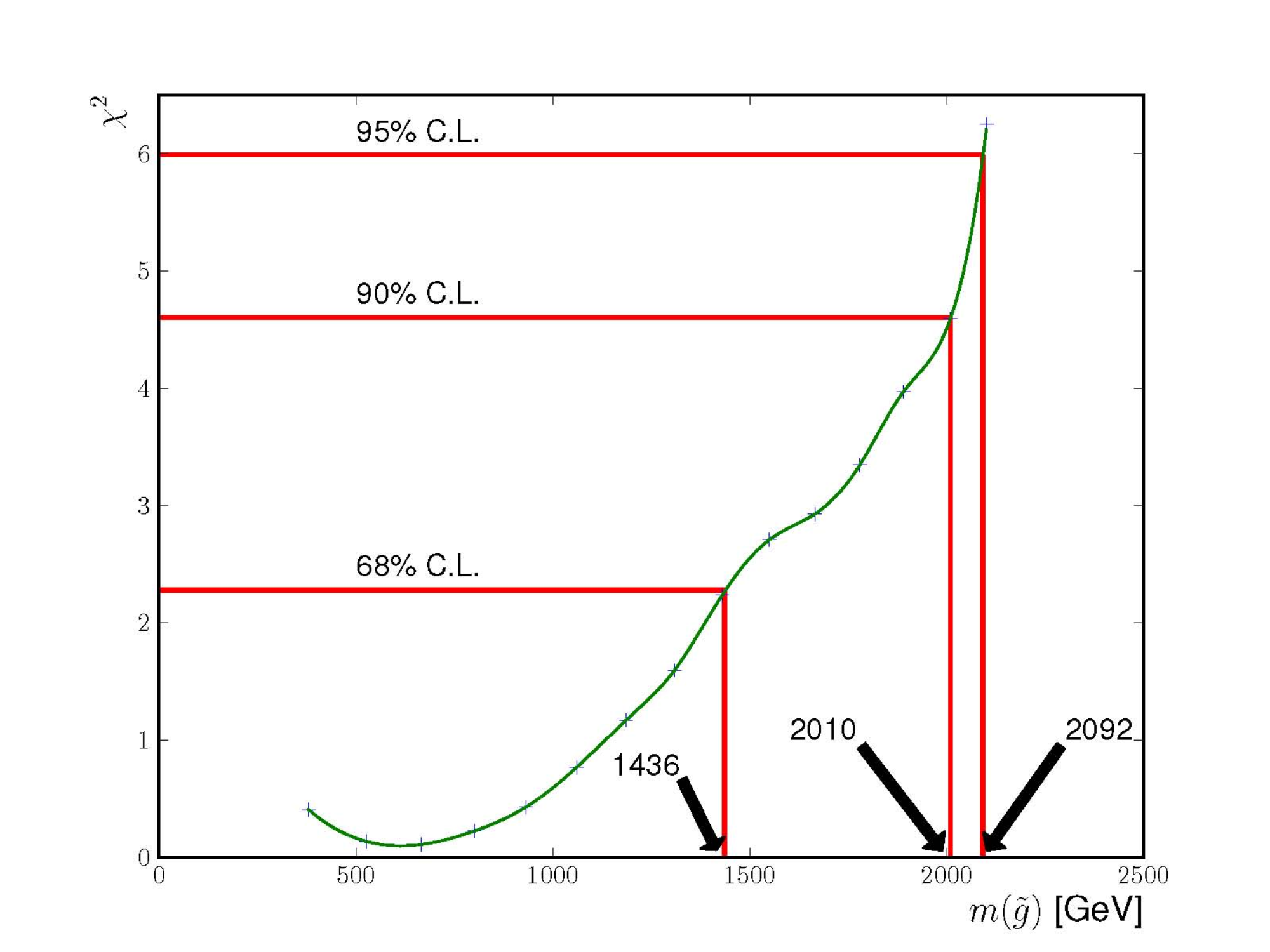}\hspace{2pc}%
\begin{minipage}[b]{30pc}\caption{\label{gluino}$\chi^2$ as a function of the gluino mass obtained by varying $M_{1/2}$ for fixed $m_{16} = 20$ TeV or 2 degrees of freedom.}
\end{minipage}
\end{figure}

We find an upper bound on the gluino mass (constrained by fitting both the bottom quark and light Higgs masses).  For $m_{16} = 20$ TeV the upper bound at 90\% CL is $m_{\tilde g} \lesssim 2$ TeV (see Fig. \ref{gluino}).  For $m_{16} = 30$ TeV the upper bound at 90\% CL increases to $m_{\tilde g} \lesssim 2.8$ TeV.  Note,  a gluino with mass $m_{\tilde g} \lesssim 1.9$ TeV should be discovered at LHC 14 with 300 fb$^{-1}$ of data at 5 $\sigma$ \cite{CMS:2013xfa}!

\begin{figure}[h]
\includegraphics[width=15pc]{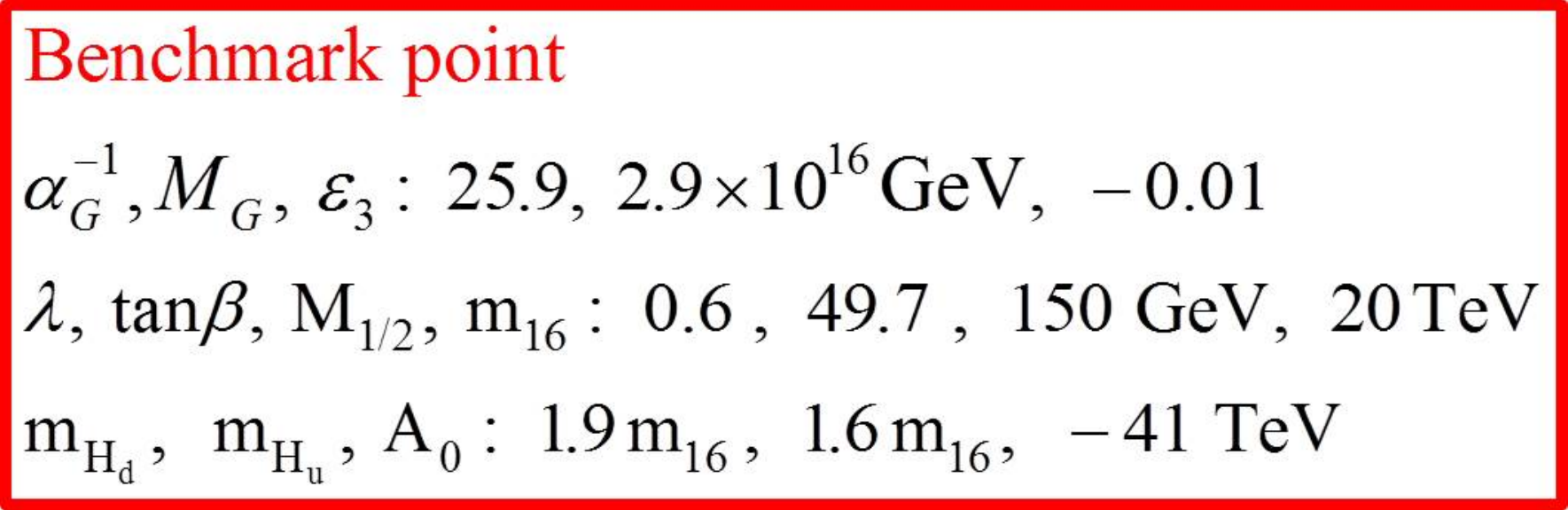}\hspace{2pc}%
\includegraphics[width=15pc]{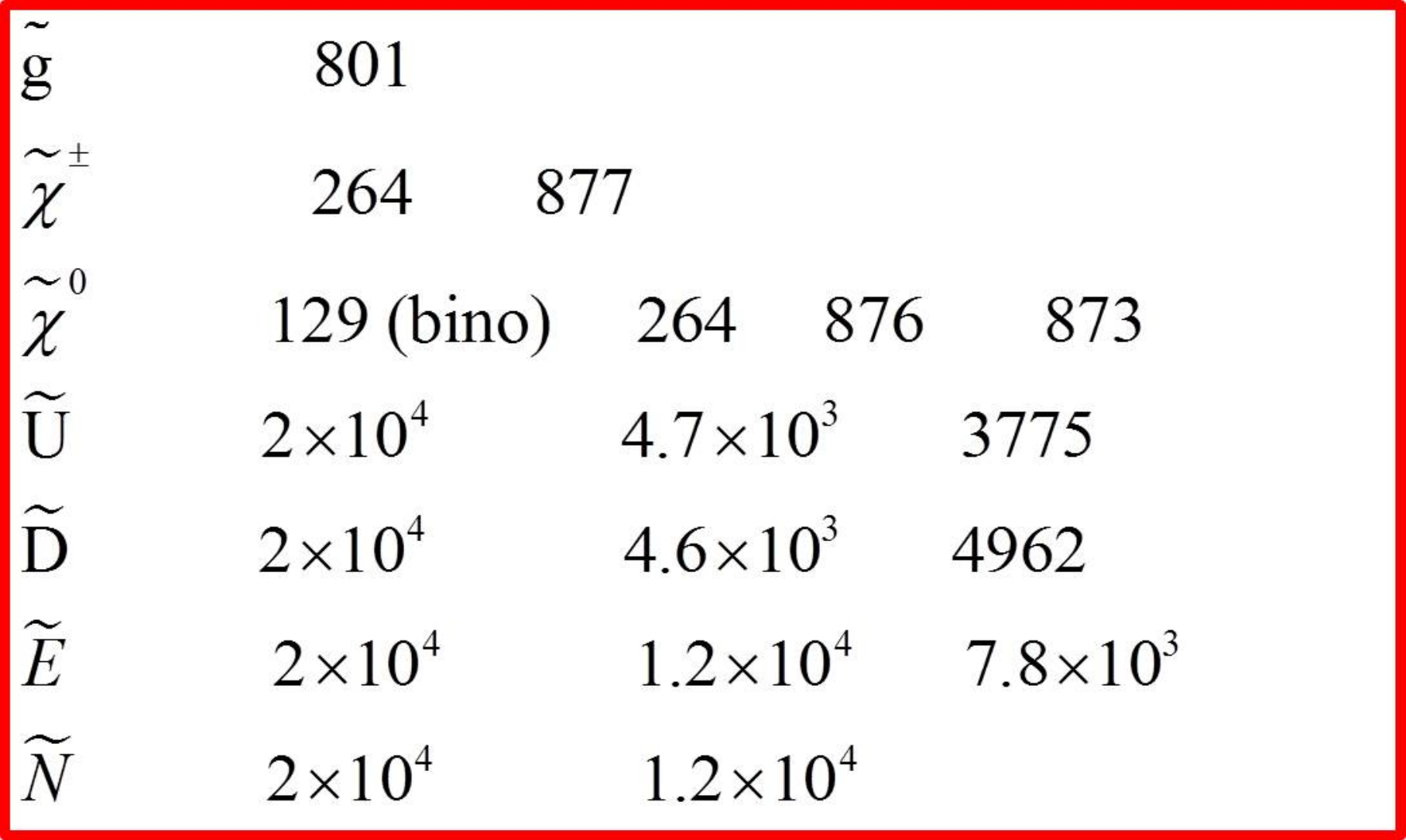}\hspace{2pc}%
\begin{minipage}[b]{30pc}\caption{\label{benchmark} (Left) The value of the 11 parameters determining the benchmark point. (Right) The SUSY particle spectrum for this benchmark point.}
\end{minipage}
\end{figure}

\begin{figure}[h]
\includegraphics[width=10pc]{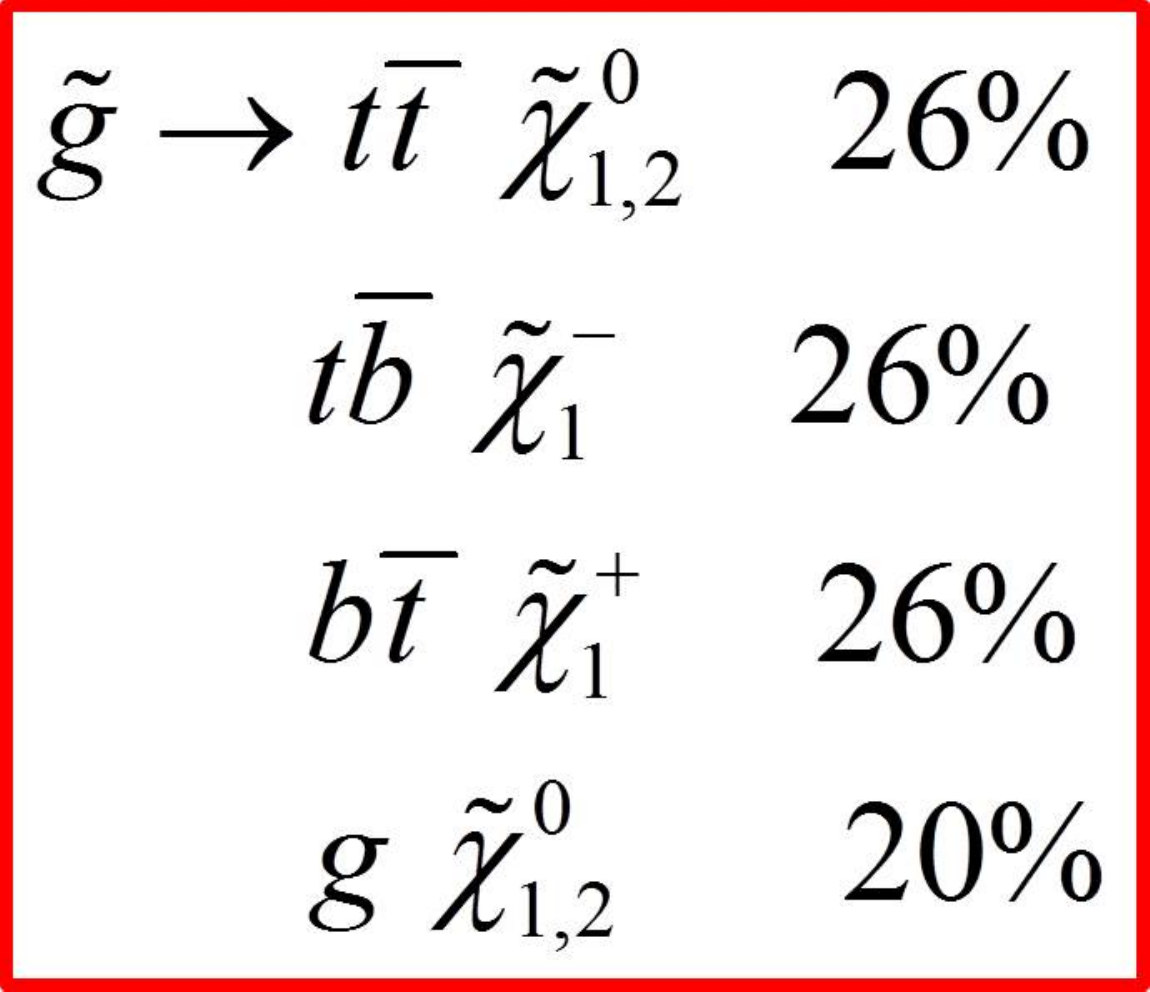}\hspace{2pc}%
\begin{minipage}[b]{30pc}\caption{\label{gluino-decay}The gluino decay branching fractions for the benchmark point obtained using SDecay.}
\end{minipage}
\end{figure}

The gluinos in our model prefer to be light, so an important question is what are the present LHC bounds on gluinos in our model?   Consider one benchmark point with the spectrum, Tables \ref{benchmark}.
The gluino decay branching fractions for this benchmark point are given in Table \ref{gluino-decay}.  Note this cannot be described by a simplified model. Hence we cannot use bounds on the gluino mass obtained using simplified models by CMS and ATLAS.
We have thus re-analyzed the data from CMS, Table \ref{CMSdata}, for 6 benchmark points with $m_{16} = 20$ TeV and different values of the gluino mass.
\begin{table}
\begin{tabular}{|c|c|c|c|}
\hline
Analysis & Luminosity & Signal Region & Reference \\
\hline
SS dilepton & 10.5 & $ N_\mathrm{jet} \geq 4,\  N_{\mathrm{b-jet}} \geq 2, $  & \cite{Chatrchyan:2012paa}\\
& & $ E_T^{\mathrm{miss}} > 120, \ H_T > 200$ & \\
\hline
$\alpha_T$ analysis (for Simplified models)  & 11.7 & $ N_\mathrm{jet} \geq 4,\  N_{\mathrm{b-jet}} = 2, \ 775 < H_T < 875$    & \cite{Chatrchyan:2013lya}\\
   (for the benchmark models)  & & $ N_\mathrm{jet} \geq 4,\  N_{\mathrm{b-jet}} \geq 2, \ 775 < H_T < 875$ &  \\
 \hline
$\Delta \phi$ analysis & 19.4 & $N_{\mathrm{b-jet}} \geq 3 ,\ E_T^{\mathrm{miss}} > 350, \ H_T > 1000$& \cite{Chatrchyan:2013wxa}\\
\hline
\end{tabular}
\caption{\label{CMSdata}The most constraining signal region for each of the analyzes studied in this work. All energies are in units of GeV and luminosity in fb$^{-1}$.}
\end{table}
We performed a detailed comparison of simplified models, in particular,  $BR(\tilde g \rightarrow t \ \bar t \ \tilde \chi^0_1) = $100\% and
$BR(\tilde g \rightarrow b \ \bar b \ \tilde \chi^0_1) = $100\%, vs. the benchmark points from our model \cite{Anandakrishnan:2013nca}. We find for the purely hadronic analyzes a 10 - 20\% less significant bound, due to the fact there are fewer b-jets as a result of the significant branching fraction, $\tilde g \rightarrow g \ \tilde \chi^0_{1,2}$.  The same sign di-lepton bounds are, on the other hand, the most significant.  The bottom line is that $m_{\tilde g} \geq 1$ TeV.

\subsection{Dark Matter}

Finally, our LSP is bino-like and thus, using microOmegas, we find it over-closes the universe.  One way to solve this problem is to include
axions.  In this case the bino can decay into an photon and axino.  While the dark matter is a linear combination of axinos and axions \cite{Baer:2008yd}.

\section{Gauge and Yukawa Unification with Non-Universal Gaugino Masses}

This part of the talk is based on the work \cite{Anandakrishnan:2013cwa} and work in progress with Archana Anandakrishnan, B. Charles Bryant and
Linda Carpenter.  We assume the following GUT scale boundary conditions, namely a universal squark
and slepton mass parameter,  $m_{16}$,  universal cubic scalar parameter, $A_0$,
``mirage" mediation gaugino masses, \begin{equation} M_i = \left(1 + \frac{g_G^2
b_i \alpha}{16 \pi^2} \log \left(\frac{M_{Pl}}{m_{16}} \right) \right) M_{1/2}
\end{equation} (where $M_{1/2}$ and $\alpha$ are free parameters and $b_i =
(33/5, 1, -3) \; {\rm for} \; i = 1, 2, 3)$.  Note, this expression is equivalent to
the gaugino masses defined in \cite{Choi:2007ka}. $\alpha$ in the above expression is related to the
$\rho$ in Ref.\cite{Lowen:2008fm} as: $\frac{1}{\rho} = \frac{\alpha}{16 \pi ^2} {\rm ln} \frac{M_{PL}}{m_{16}}$.
We consider two different cases
for non-universal Higgs masses [NUHM] with ``just so'' Higgs splitting
\begin{equation}  m_{H_{u (d)}}^2 = m_{10}^2  - (+) 2 D \end{equation} with universal squark and slepton masses, $m_{16}$,  or, D-term
Higgs splitting, where, in addition, squark and slepton masses are given by
\begin{equation} m_a^2 =  m_{16}^2 + Q_a D, \;\; \{ Q_a = +1,  \{ Q, \bar u,
\bar e \}; -3, \{ L, \bar d \} \} \end{equation} with the U(1) D-term, $D$, and
SU(5) invariant charges, $Q_a$.   Note, we take $\mu, \, M_{1/2} < 0$. Thus for
$\alpha \ge 4 $ we have $M_3 > 0, M_1, M_2 < 0$.

\begin{table}
\renewcommand{\arraystretch}{1.2}
\scalebox{0.83}{
\begin{tabular}{|l||c|}
\hline
Sector &  Third Family Analysis  \\
\hline
gauge             & $\alpha_G$, $M_G$, $\epsilon_3$          \\
SUSY (GUT scale)  & $m_{16}$, $M_{1/2}$, $\alpha$, $A_0$, $m_{10}$, $D$
\\
textures          & $\lambda$                                         \\
SUSY (EW scale)   & $\tan \beta$, $\mu$                               \\
\hline
Total \#          &                                              12 \\
\hline
\end{tabular}
}
\caption{\footnotesize The 12 parameters defining the model.}
\label{tab:parameters}
\end{table}

\begin{figure}[h]
\includegraphics[width=20pc]{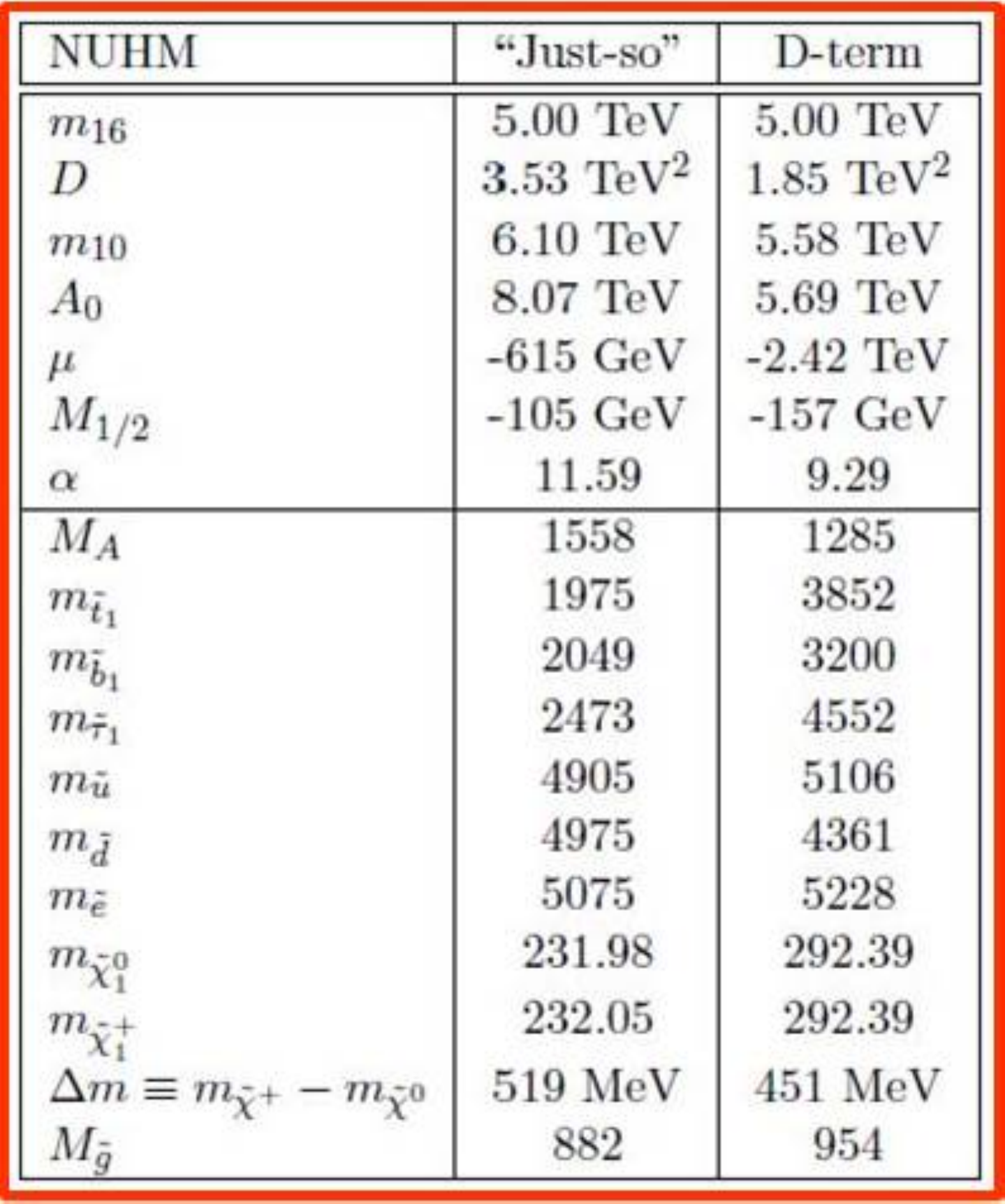}\hspace{2pc}%
\begin{minipage}[b]{30pc}\caption{\label{mirage}The SUSY spectrum for two benchmark points.}
\end{minipage}
\end{figure}

There are 12 parameters and 11 observables, thus we require $\chi^2 \ll 1$.  Nevertheless, since Yukawa unification severely
constrains the SUSY breaking sector of the theory we are confident that the SUSY spectrum is robust.  Two benchmark points are given in Table \ref{mirage}.
Note, the parameter $\alpha$ which determines the ratio of anomaly mediated and gravity mediated SUSY breaking is large.  Thus
the spectrum is similar to that of pure anomaly mediation with an almost degenerate neutralino and chargino; both predominantly wino-like.
The neutralino and chargino masses in Table \ref{mirage} are tree level running masses and the factor $\Delta m$ includes the one loop correction
to their masses.   Note the splitting is of order 500 MeV.  As a result the chargino decays predominantly into the neutralino and a single pion.
The gluino decay branching fractions for the benchmark point obtained using SDecay is
\begin{eqnarray} \tilde g \rightarrow & \{ 63\% \;\; \tilde \chi^0 \ g, \;  {\rm and \ the \ rest \ to}  \; \tilde \chi^+ \ b \ \bar t, \;  \tilde \chi^- \ t \ \bar b \} &  {\rm Just-so \ splitting}\\
\tilde g \rightarrow & \{ 56\% \;\; \tilde \chi^+ \ b \ \bar t, \;  \tilde \chi^- \ t \ \bar b;  \; 17\% \;\; \tilde \chi^0 \ t \ \bar t; \; 10\% \;\; \tilde \chi^0 \ b \ \bar b, \; {\rm and \ the \ rest \ to \ light \ quarks} \} &  {\rm D-term \ splitting} \nonumber
\end{eqnarray}
We are now studying the LHC bounds on the sparticle masses in this model.

\subsection{Dark Matter}
In this model the dark matter candidate is predominantly wino-like.  Therefore, using microOmegas we find, assuming thermal dark matter, that the universe is under-closed.   This problem can be avoided if winos are produced non-thermally or with another source of dark matter, such as axions.

\section{3 Family Model}

The previous results depended solely on SO(10) Yukawa unification for the third family.   We now consider a complete three family SO(10) model for fermion masses and mixing, including neutrinos \cite{Dermisek:2005ij,Dermisek:2006dc,Albrecht:2007ii}.   The model also includes a
$D_3 \times [U(1) \times \mathbb{Z}_2 \times \mathbb{Z}_3]$ family symmetry which is necessary to
obtain a predictive theory of fermion masses by reducing the number of arbitrary parameters in the Yukawa matrices.
In the rest of this talk we will consider the new results due to the three family analysis.  We shall consider the superpotential generating the effective fermion Yukawa couplings.  We then perform a global $\chi^2$ analysis, including precision electroweak data which now includes both neutral and charged fermion masses and mixing angles.

\begin{figure}[h]
\includegraphics[width=20pc]{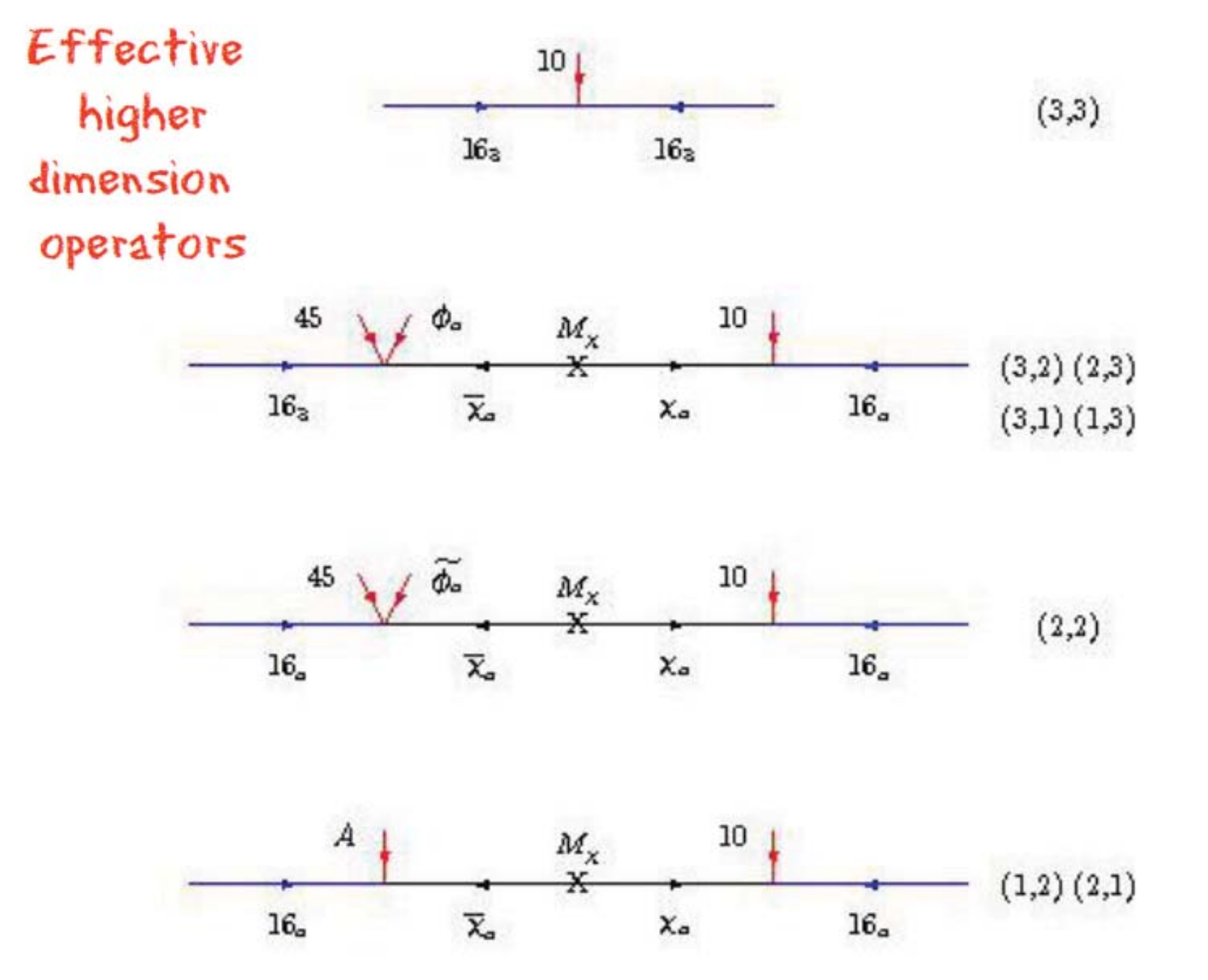}\hspace{2pc}%
\begin{minipage}[b]{20pc}\caption{\label{massoperators}The effective fermion mass operators obtained after integrating out the Froggatt-Nielsen massive states.}
\end{minipage}
\end{figure}

\begin{figure}[h]
\includegraphics[width=14pc]{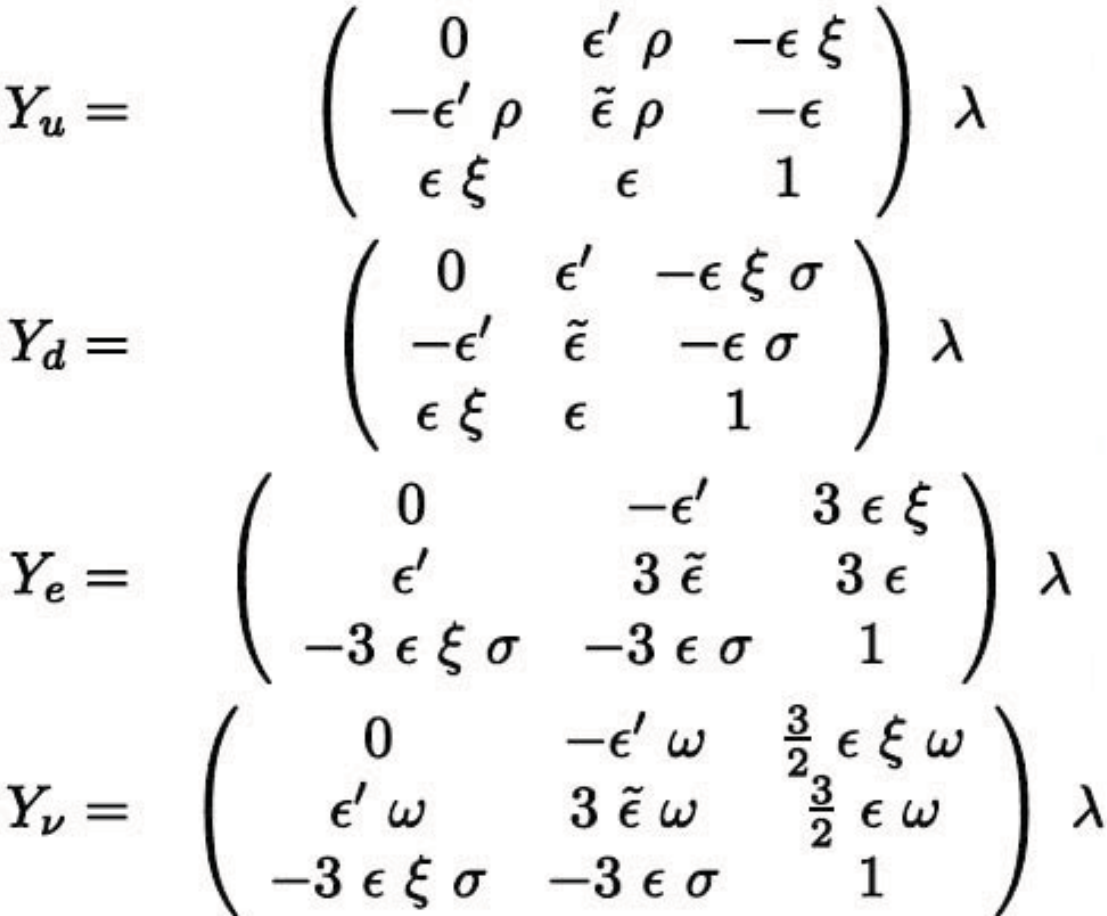}\hspace{2pc}%
\begin{minipage}[b]{14pc}\caption{\label{yukawas}The Yukawa matrices obtained from the effective fermion mass operators after taking into account the flavon VEVs.}
\end{minipage}
\end{figure}

The superspace potential for the charged fermion sector of this model is
given by:
\begin{eqnarray} W_{ch. fermions} = & 16_3 \ 10 \ 16_3 +  16_a \ 10 \ \chi_a & \label{Wchf}
\\ & +  \bar \chi_a \ ( M_{\chi} \ \chi_a + \ 45 \ \frac{\phi_a}{\hat M} \ 16_3 \ + \ 45 \ \frac{\tilde
\phi_a}{\hat M} \  16_a + {\bf A} \ 16_a ) & \nonumber
\end{eqnarray}
where $45$ is an $SO(10)$ adjoint field which is assumed to obtain
a VEV in the B -- L direction; and $M$ is a linear combination of an
$SO(10)$ singlet and adjoint. Its VEV $M_0 ( 1 + \a X + \b Y)$ gives mass
to Froggatt-Nielsen states. Here $X$ and $Y$ are elements of the Lie
algebra of $SO(10)$ with $X$ in the direction of the $U(1)$ which
commutes with $SU(5)$ and $Y$ the standard weak hypercharge; and $ \a $ ,
$ \b $ are arbitrary constants which are fit to the data.

\begin{equation}
\p, \nonumber
\quad \ptil, \nonumber
\quad A, \nonumber
\end{equation}
are $SO(10)$ singlet 'flavon' fields, and
\begin{equation}
\bar \chi_a, \nonumber
 \quad \chi_a
\nonumber
\end{equation}
are a pair of Froggatt-Nielsen states transforming as a $\overline {16}$ and $16$ under $SO(10)$.   The 'flavon' fields are {\em assumed} to obtain VEVs of the form
\begin{equation}  \langle \p \rangle = \left( \begin{array}{c} \phi_1 \\ \phi_2 \end{array} \right), \;\;
\langle \ptil \rangle = \left( \begin{array}{c} 0 \\ \tilde \phi_2 \end{array} \right). \end{equation}
After integrating out the Froggatt-Nielsen states one obtains the effective fermion mass operators in Fig. \ref{massoperators}.
We then obtain the Yukawa matrices for up and down quarks, charged leptons and neutrinos given in Fig. \ref{yukawas}.  These matrices contain 7
real parameters and 4 arbitrary phases.
Note,  the superpotential (Eqn. \ref{Wchf}) has many arbitrary parameters.   However, at the end of the day the effective Yukawa matrices have many fewer parameters.
This is good, because we then obtain a very predictive theory.   Also, the quark mass matrices accommodate the Georgi-Jarlskog mechanism, such that
$m_\mu/m_e \approx 9 m_s/m_d$.

We then add 3 real Majorana mass parameters for the neutrino see-saw mechanism. The anti-neutrinos get GUT scale masses by mixing with
three $SO(10)$ singlets $\{ N_a, \ a = 1,2;
\;\; N_3 \}$ transforming as a $D_3$ doublet and singlet respectively. The full superpotential is given by $W =
W_{ch. fermions} + W_{neutrino}$ with
\begin{eqnarray} \label{eq:WneutrinoD3} W_{neutrino} = & \overline{16} \left(\lambda_2 \ N_a \ 16_a \ + \ \lambda_3 \ N_3 \ 16_3 \right) & \\
& +  \;\; \frac{1}{2} \left(S_{a} \ N_a \ N_a \;\; + \;\; S_3 \ N_3 \ N_3\right).   & \nonumber
\end{eqnarray}
We assume $\overline{16}$ obtains a VEV, $v_{16}$, in the right-handed neutrino direction, and $\langle S_{a}
\rangle = M_a$ for $a = 1,2$ and $\langle S_3 \rangle = M_3$.  The effective neutrino mass terms are given by
\begin{equation} W =  \nu \ m_\nu \ \bar \nu + \bar \nu \ V \ N +
\frac{1}{2} \ N \ M_N \ N \end{equation} with
\begin{equation} V = v_{16} \ \left(
\begin{array}{ccc} 0 &  \lambda_2 & 0 \\
\lambda_2 & 0 & 0 \\ 0 & 0 &  \lambda_3 \end{array} \right), \; M_N = diag( M_1,\ M_2,\ M_3)  \end{equation}  all assumed
to be real.  Finally, upon integrating out the heavy Majorana neutrinos we obtain the $3 \times 3$ Majorana mass matrix for the light
neutrinos in the lepton flavor basis given by
\begin{equation}
{\cal M} =   U_e^T \ m_\nu  \ M_R^{-1} \ m_\nu^T  \ U_e ,
\end{equation}
where the effective right-handed neutrino Majorana mass matrix is given by:
\begin{equation}
M_R =  V \ M_N^{-1}  \ V^T  \  \equiv \  {\rm diag} ( M_{R_1}, M_{R_2}, M_{R_3} ),
\end{equation}
with \begin{eqnarray} M_{R_1} = (\lambda_2 \ v_{16})^2/M_2, \quad  M_{R_2} = (\lambda_2 \ v_{16})^2/M_1, \quad  M_{R_3} =
(\lambda_3 \ v_{16})^2/M_3 . \label{eq:rhmass} \end{eqnarray}

\section{Global $\chi^2$ analysis}

Just in the fermion mass sector we can see that the theory is very predictive.   We have 15 charged fermion and 5 neutrino low energy observables given
in terms of 11 arbitrary Yukawa parameters and 3 Majorana mass parameters.  Hence there are 6 degrees of freedom in this sector of the theory.   However
in order to include the complete MSSM sector we perform the global $\chi^2$ analysis with 24 arbitrary parameters at the GUT scale given in Table \ref{paras}.
Note, this is to be compared to the 27 arbitrary parameters in the Standard Model or the 32 parameters in the CMSSM.
\begin{table}[h]
\caption{\label{paras}Parameters entering the global $\chi^2$ analysis.}
\begin{tabular}{|l|c|c|}
\hline
$Sector$&\#&$Parameters$\cr
\hline
$gauge$&$3$  &$\alpha_G, \ M_G, \ \epsilon_3$\cr
$SUSY \ (GUT scale)$&$5$  &$m_{16}, \ M_{1/2}, \ A_0, \ m_{H_u}, \ m_{H_d}$\cr
$textures$&$11$  &$\lambda, \ \epsilon, \ \epsilon', \ \rho, \ \sigma, \ \tilde \epsilon, \ \xi$\cr
$neutrino$&$3$  &$M_{R_1}, \ M_{R_2}, \ M_{R_3}$\cr
$SUSY \ (EW scale)$&$2$  &$\mu, \ \tan\beta$\cr
\hline
\end{tabular}
\end{table}

In this work we have decided to extend the analysis of Albrecht et al. to values of $m_{16} \ge 10$ TeV,  including more low energy observables such as the light Higgs mass,  the neutrino mixing angle $\theta_{13}$ and lower bounds on the gluino and squark masses coming from recent data.
We perform a three family global $\chi^2$ analysis.  We are using the code, \maton, developed by Radovan Dermisek to renormalize the parameters in the theory from the GUT scale to the weak scale,  perform electroweak symmetry breaking and calculate squark, slepton, gaugino masses, as well as quark and lepton masses and mixing angles. We also use the Higgs code of Pietro Slavich (suitably revised for our particular scalar spectrum) to calculate the light Higgs mass and SUSY\_Flavor\_v2.0 \cite{Crivellin:2012jv} to evaluate flavor violating B decays.

There are 24 arbitrary parameters defined mostly at the GUT scale and run down to the weak scale where the $\chi^2$ function is evaluated.  However
the value of $m_{16}$ has been kept fixed in our analysis, so that we can see the dependence of $\chi^2$ on this input parameter.   Thus with
23 arbitrary parameters we fit 36 observables, giving 13 degrees of freedom.  The $\chi^2$ function has been minimized using the CERN package, MINUIT.

\vspace{.1cm}

Initial parameters for benchmark point with $m_{16} = 20$ TeV (see Table \ref{t:fit20tev}). \\
(1/$\alpha_G, \, M_G, \, \epsilon_3$) = ($25.90, \, 3.13 \times 10^{16}$ GeV, $\, -1.45$ \%), \\
($\lambda, \, \lambda \epsilon, \, \sigma, \, \lambda \tilde \epsilon, \, \rho, \, \lambda \epsilon', \, \lambda \epsilon \xi$) = ($ 0.60, \, 0.031, \, 1.14, \, 0.0049, \,
0.070, \, -0.0019, \, 0.0038  $),\\
($\Phi_\sigma, \, \Phi_{\tilde \epsilon}, \, \Phi_\rho, \, \Phi_\xi$) =  ($0.533, \, 0.548, \, 3.936, \, 3.508$) rad, \\
($m_{16}, \, M_{1/2}, \, A_0, \, \mu(M_Z)$) = ($20000,\, 168, \, -41087, \, 1163.25$) GeV,\\
($(m_{H_d}/m_{16})^2, \, (m_{H_u}/m_{16})^2, \, \tan\beta$) = ($1.85,  \, 1.61, \, 49.82$) \\
($M_{R_3}, \, M_{R_2}, \, M_{R_1}$) = ($3.2 \times 10^{13}$ GeV, $\, 6.1 \times 10^{11} $ GeV, $\, 0.9 \times 10^{10} $ GeV)
\vspace{.1cm}

The fit is quite good with $\chi^2/d.o.f. = 2$.   However, note that we have not taken into account correlations in the data, so we will just use $\chi^2$ as a indicator of the rough goodness of the fit.
\vspace{5mm}
\begin{table}
\centering
\caption{Benchmark point with $ {\boldsymbol{m_{16} = 20 \; {\rm TeV}}}$. \label{t:fit20tev}}
\begin{tabular}{|l|l|l|l|l|}
\hline
Observable  &  Fit value  &  Exp value  &  Pull & Sigma  \\
\hline
$M_Z$ &              91.1876         &  91.1876         &  0.0000          &  0.4559          \\
$M_W$ &              80.5452         &  80.3850         &  0.3982          &  0.4022          \\
$1/\alpha_{em}$ &    137.0725        &  137.0360        &  0.0533          &  0.6852          \\
$G_{\mu} \times 10^5$ & 1.1713          &  1.1664          &  0.4250          &  0.0117          \\
$\alpha_3$ &         0.1184          &  0.1184          &  0.0467          &  0.0009          \\
\hline
$M_t$ &              174.0184        &  173.5000        &  0.3916          &  1.3238          \\
$m_b(m_b)$ &         4.1849          &  4.1800          &  0.1334          &  0.0366          \\
$M_{\tau}$ &         1.7755          &  1.7768          &  0.1462          &  0.0089          \\
\hline
$m_c(m_c)$ &         1.2547          &  1.2750          &  0.7876          &  0.0258          \\
$m_s$ &              0.0964          &  0.0950          &  0.2807          &  0.0050          \\
$m_d/m_s$  &         0.0692          &  0.0526          &  2.9891          &  0.0055          \\
$1/Q^2$ &            0.0018          &  0.0019          &  0.4749          &  0.0001          \\
$M_{\mu}$ &          0.1056          &  0.1057          &  0.1049          &  0.0005          \\
$M_e \times 10^4$ &  5.1122          &  5.1100          &  0.0862          &  0.0255          \\
\hline
$|V_{us}|$ &         0.2243          &  0.2252          &  0.5964          &  0.0014          \\
$|V_{cb}|$ &         0.0415          &  0.0406          &  0.4511          &  0.0020          \\
$|V_{ub}| \times 10^3$ & 3.2023          &  3.7700          &  0.6678          &  0.8502          \\
$|V_{td}| \times 10^3$ & 8.9819          &  8.4000          &  0.9675          &  0.6015          \\
$|V_{ts}|$ &         0.0407          &  0.0429          &  0.8518          &  0.0026          \\
$\text{sin} 2\beta$ & 0.6304          &  0.6790          &  2.3959          &  0.0203          \\
\hline
$\epsilon_K$ &       0.0023          &  0.0022          &  0.3823          &  0.0002          \\
$\Delta M_{Bs}/\Delta M_{Bd}$ & 39.4933         &  35.0600         &  0.6311          &  7.0246          \\
$\Delta M_{Bd} \times 10^{13}$ & 3.9432          &  3.3370          &  0.9072          &  0.6682          \\
\hline
$m^2_{21} \times 10^5$ &   7.5126          &  7.5450          &  0.0593          &  0.5463          \\
$m^2_{31} \times 10^3$ &  2.4828          &  2.4800          &  0.0135          &  0.2104          \\
$\text{sin}^2 \theta_{12}$ & 0.2949          &  0.3050          &  0.2880          &  0.0350          \\
$\text{sin}^2 \theta_{23}$ & 0.5156          &  0.5050          &  0.0640          &  0.1650          \\
$\text{sin}^2 \theta_{13}$ & 0.0131          &  0.0230          &  1.4134          &  0.0070          \\
\hline
$M_h$ &              124.07          &  125.30          &  0.4010          &  3.0676          \\
\hline
$BR(B \rightarrow X_s \gamma) \times 10^4$ & 3.4444          &  3.4300          &  0.0088          &  1.6374          \\
$BR(B_s \rightarrow \mu^+ \mu^-) \times 10^9$ & 1.6210          &  3.2000          &  0.9682          &  1.6308          \\
$BR(B_d \rightarrow \mu^+ \mu^-) \times 10^{10}$ & 1.0231          &  8.1000          &  0.0000          &  5.2559          \\
$BR(B \rightarrow \tau \nu) \times 10^5$ & 6.3855          &  16.6000         &  1.1436          &  8.9320          \\
$BR(B \rightarrow K^*\mu^+ \mu^-)$(low) $\times 10^8$& 5.1468          &  19.7000         &  1.2123          &  12.0051         \\
$BR(B \rightarrow K^*\mu^+ \mu^-)$(high) $\times 10^8$ & 7.7469          &  12.0000         &  0.5839          &  7.2835          \\
$q_0^2(B \rightarrow K^* \mu^+ \mu^-)$ & 4.5168          &  4.9000          &  0.2945          &  1.3009          \\
\hline
\multicolumn{3}{|l}{Total $\chi^2$}  &  \textbf{26.5812}&  \\\hline
\end{tabular}
\end{table}

In Table \ref{spectrum} we see that the value of $\chi^2$ decreases as $m_{16}$ increases,  but our analysis shows that the
$m_{16} \sim 20$ TeV minimizes $\chi^2$, i.e. we have found that $\chi^2$ slowly increases for $m_{16} > 20$ TeV.  Note, we are able to fit the neutrino masses and mixing angles quite well.  The two large mixing angles are due to the hierarchy of right-handed neutrino masses.  The biggest discrepancy is for the angle $\theta_{13}$.  We obtain a value which is closer to 6$^\circ$, rather than the observed value of order 9$^\circ$. In Table \ref{predictions} we present results for lepton flavor and CP violation.

\begin{table}
\centering
\begin{tabular}{|l|c|c|}
\hline
$m_{16}  $  &   20 TeV & 30 TeV \\
%$A_0$  &  -41.1 TeV &  -61.6 TeV   \\
%$\mu  $  & 1163 &  1647    \\
%$M_{1/2}$ &  168  & 162   \\
\hline
$\chi^2$ &  26.58 &  29.48  \\
\hline
$M_A$ & 1651 &   2036  \\
$m_{\tilde t_1}$ &  3975 &  5914   \\
$m_{\tilde b_1}$  &  5194 &  7660   \\
$m_{\tilde \tau_1}$  &  7994 &  11620   \\
$m_{\tilde\chi^0_1}$   &  137 &   167     \\
$m_{\tilde\chi^+_1}$  & 279 & 351     \\
$M_{\tilde g}$   & 851 &   1004    \\
\hline
$\chi^2/{\rm dof}$ & 2 & 2.2 \\
\hline
\end{tabular}
\caption{\footnotesize SUSY Spectrum corresponding to two benchmark points. The first two generation scalars have mass of the order of $m_{16}$.} \label{spectrum}
\end{table}

\begin{table}
 \centering
\begin{tabular}{|l|l|l|l|l|l|l|}
\hline
& Current Limit & 10 TeV & 15 Te V & 20 TeV & 25 TeV & 30 TeV \\
\hline
e EDM $\times 10^{28}$ & $< 10.5 \ e\ cm$ & $-0.224$ & $-0.0408 $ & $-0.0173 $ & $-0.0113 $ & $-0.0084 $\\
$\mu$ EDM $\times 10^{28}$& $(-0.1 \pm 0.9) \times 10^{9}\ e\ cm$ & $34.6 $ & $6.23 $& $3.04 $ & $1.77 $ & $1.20$\\
$\tau$ EDM $\times 10^{28}$ & $-0.220 - 0.45 \times 10^{12}\ e\ cm$ & $-2.09 $ & $-0.394 $& $-0.185 $ & $-0.109 $ & $-0.0732 $\\
\hline
BR$(\mu \rightarrow e \gamma) \times 10^{12}$ & $< 2.4$ & $5.09 $ &$1.23 $ & $0.211 $ & $0.0937 $ & $0.0447 $\\
BR$(\tau \rightarrow e \gamma) \times 10^{12}$  & $< 3.3 \times 10^{4}$ & $58.8 $ &$13.9 $ & $2.40 $ & $1.04 $ &$0.502 $ \\
BR$(\tau \rightarrow \mu \gamma) \times 10^{8}$  & $< 4.4 $ &$1.75 $ &$0.498 $ & $0.0837 $& $0.0385 $& $0.0182 $\\
\hline
sin $\delta$ & & -0.60 & -0.87 & -0.27 & -0.42 & -0.53 \\
\hline
\end{tabular}
\caption{\label{predictions} \footnotesize Predictions from the full three family analysis. The dipole moments and branching ratios were calculated using \susyflavor.}
\end{table}

\section{Conclusion and plans for the future}

We have presented an analysis of a theory satisfying Yukawa unification and large $\tan\beta$.  The results are encouraging.
We find SO(10) Yukawa unification is still alive after LHC 7, 8!
Some of the good features are -
\begin{itemize}
\item  gauge coupling unification is satisfied
\item  the Higgs mass is of order 125 GeV and Standard Model-like
\item  there is an inverted scalar mass hierarchy
\item  for universal gaugino masses we find $m_{\tilde g} \geq 1$ TeV and $m_{\tilde g} \leq 3$ TeV for $m_{16} \leq 30$ TeV
\item  for non-universal gaugino masses, the lightest chargino and neutralino are almost degenerate.
\end{itemize}
Finally, in collaboration with Anandakrishnan, Bryant and Carpenter, we are continuing to analyze the LHC phenomenology of models with effective ``mirage" mediation.

\begin{theacknowledgments}
I wish to thank the organizers of both the VIIth International Conference on Interconnections between Particle Physics and Cosmology
and CETUP* for financial support and the wonderful conference/workshop.  I also acknowledge
partial support from DOE grant DOE/ER/01545-900.
\end{theacknowledgments}

\bibliographystyle{aipproc}   % if natbib is available

\bibliography{bibliography}

\bibliographystyle{utphys}

\end{document}